\documentclass[aps,pre,showpacs,twocolumn,superscriptaddress,floatfix]{revtex4}

\usepackage{graphicx}
\usepackage{amssymb}
\usepackage{amsmath}

\begin{document}

\title{Extreme value statistics and return intervals in long-range correlated
uniform deviates}

\author{N. R. Moloney}
\email{nmoloney@phas.ucalgary.ca}
\author{J. Davidsen}
\email{davidsen@phas.ucalgary.ca}

\affiliation{Department of Physics and Astronomy, University of Calgary, 2500
  University Drive NW, Calgary, AB T2N 1N4, Canada.}

\date{\today}

\begin{abstract}
We study extremal statistics and return intervals in \emph{stationary}
long-range correlated sequences for which the underlying probability density
function is bounded and uniform.  The extremal statistics we consider
(e.g. maximum relative to minimum) are such that the reference point from
which the maximum is measured is itself a random quantity. We analytically
calculate the limiting distributions for independent and identically
distributed random variables, and use these as a reference point for
correlated cases. The distributions are different from that of the maximum
itself (i.e. a Weibull distribution), reflecting the fact that the
distribution of the reference point either dominates over or convolves with
the distribution of the maximum. The functional form of the limiting
distributions is unaffected by correlations, although the convergence is
slower. We show that our findings can be directly generalized to a wide class
of stochastic processes.  We also analyze return interval distributions, and
compare them to recent conjectures of their functional form.
\end{abstract}

\maketitle

\section{Introduction}

Interest in the extreme behavior of complex systems has been
growing, with examples including the estimation of DNA replication
times~\cite{BechhoeferMarshall:2007}, extreme paths on random
trees~\cite{MajumdarKrapivsky:2000,BenNaimKrapivskyMajumdar:2001}
and their applications to computer
science~\cite{MajumdarKrapivsky:2002}, extreme eigenvalues in
random matrices and low-lying states in disordered
systems~\cite{BouchaudMezard:1997,BiroliBouchaudPotters:2007}, and
extreme values in multifractal
processes~\cite{MuzyBacryKozhemyak:2006}. In particular, much
attention has been paid to extreme value statistics in time
series. This is especially relevant in the context of disasters and 
hazard assessment~\cite{BundeKroppSchellnhuber:2002}, flood prediction
being a notable example~\cite{BattjesGerritsen:2002}. In regard to
this, results exist in the mathematical
literature~\cite{LeadbetterLindgrenRootzen:1983} which extend the
range of extreme limit distributions based on identically and
independently distributed (iid) random variables to a wide class
of dependent time series, see~\cite{LeadbetterRootzen:1988} for a
review. Yet, the exact extremal properties of many time series 
exhibiting long-range correlations are far from being fully 
understood. Examples of such time series in nature include 
crackling noise~\cite{sethna01}, water levels in
rivers~\cite{hurst51}, temperature fluctuations in
oceans~\cite{MonettiHavlinBunde:2002}, or climatological temperature
records~\cite{pelletier97,huybers06}. Much insight into the
extremes of these time series can be gained by studying so-called
$1/f^{\alpha}$
signals~\cite{MajumdarComtet:2004,GyorgyiETAL:2007,BurkhardtETAL:2007}
(in reference to the power-law decay of the power spectrum), which
capture many of the essential features of real-world long-range
correlated time series.

Aside from the distribution of extremes, another very useful and practical
indicator for hazard assessment based on time series is the distribution of
return intervals between successive threshold-crossing events.  For
uncorrelated time series, the Poisson process gives rise to the well-known
exponential return interval distribution~\cite{CoxIsham:1980}. For long-range
correlated series encountered in nature, however, the distribution is no
longer exponential and a number of distributions have been put forward,
depending on the data set, such as gamma distributions and power-laws with
stretched exponential
tails~\cite{Corral:2004,DavidsenStanchitsDresen:2007,baiesi06m,BundeETAL:2005,VazquezETAL:2006}.
Artificially generated $1/f^{\alpha}$ signals have been studied in
detail~\cite{AltmannKantz:2005,EichnerETAL:2007} and some progress has been
made towards a theoretical understanding of return intervals in long-range
correlated
series~\cite{Corral:2005,SornetteUtkinSaichev:2008,SanthanamKantz:2008}, but
the overall picture is far from complete.

Here, we focus on the extremal properties and return intervals of stationary
long-range correlated series for which the underlying probability density
function (PDF) is uniform over a given finite interval, i.e. the $X_i$ that
make up the series $\{X_i\}_{i=1 \dots n}$ are uniformly distributed. These
processes are of general interest and have been neglected in the past. For iid
random variables, a single parameter family of distributions describes the
possible limiting distribution for the maximum of a
sample~\cite{deHaanFerreira:2006}. This family is traditionally broken up into
three qualitatively different distributions: Fr\'echet, Gumbel and
Weibull. Provided certain conditions are met, a given underlying PDF for the
iid random variables will fall within the domain of attraction of one of these
three extreme value distributions. Previous studies have concentrated on time
series with underlying PDF that belong to the Gumbel or Fr\'echet domains of
attraction.  Random variables distributed uniformly on $[0,1]$, however,
belong to the Weibull domain of attraction~\footnote{We note in passing that
the upper-boundedness of the distribution is not a sufficient condition for
the maxima to be Weibull distributed.}. In fact, the same is true for a series
of stationary long-range correlated uniform deviates, provided certain mixing
criteria are
met~\cite{LeadbetterLindgrenRootzen:1983,LeadbetterRootzen:1988}. However, we
show in this paper that a variety of extremal quantities converge to very
different limiting distributions. When the maximum is measured relative to a
quantity that is itself a random variable, the extremal distribution need no
longer be Gumbel, Fr\'{e}chet or Weibull. In this paper we consider the
following three extremal quantities: (a) maximum relative to the average (over
the $X_i$), introduced in the context of
interfaces~\cite{RaychaudhuriETAL:2001}, (b) maximum relative to the initial
value~\cite{BurkhardtETAL:2007,SabhapanditMajumdar:2007}, which is a natural
measure in the context of time series (e.g. the maximum increase of a stock
from its starting price), (c) maximum relative to minimum, which is a measure
of the full range of values encountered (e.g. as measured in auroral
indices~\cite{HnatETAL:2002b}).
 
The structure of the paper is as follows: In Section~\ref{S:iid} we present
analytic results for extremal distributions for iid random variables and we
briefly discuss how our results may be generalised to other distributions
belonging to the Weibull class, e.g. the beta distribution.  In
Section~\ref{S:stationary}, we then compare the iid results to the respective
distributions in stationary long-range correlated series. In
Section~\ref{S:returns} we present the corresponding return interval
distributions and examine their asymptotic behavior. We conclude the paper in
Section~\ref{S:conclusion}. In an Appendix, we discuss some of the features of
the Schreiber-Schmitz
algorithm~\cite{SchreiberSchmitz:1996,SchreiberSchmitz:2000} for generating
long-range correlated series with a specified underlying PDF.

\section{Extremal statistics for iid uniform deviates\label{S:iid}}

Consider a set of $n$ iid random variables $\{X_1,\ldots,X_n\}$ drawn from the
(cumulative) distribution function $F(z)$. The distribution of the maximum
$M_n$ is
\begin{align}
  \text{Pr}(M_n \le z) &= \text{Pr}(X_1 \le z, \ldots, X_n \le z) \notag \\
       &= F^n(z)
\end{align}
A fundamental result of extreme value theory is that if $F^n(a_n z
+ b_n)$ converges to a non-degenerate limiting distribution as
$n\to\infty$, where $a_n$ and $b_n$ are scale and location
parameters, respectively, that effect a linear rescaling, the
distribution can be one of only three types. The eventual limiting
distribution is determined by the asymptotic behavior of the
underlying $F(z)$~\cite{deHaanFerreira:2006}. For the specific
example of $n$ iid uniform deviates with probability density $p(z)
= 1$ for $0 \le z \le 1$, the choice $a_n = 1/n, b_n = 1$ gives
\begin{align}
  \lim_{n\to\infty}F^n(z/n + 1) &= \lim_{n\to\infty} (z/n+1)^n \notag \\
    &= e^z\label{E:Weibull},
\end{align}
which is an example of a Weibull distribution. While the choice of $a_n$ and
$b_n$ is not unique, the particular choice does not influence the functional
form of the limiting distribution according to the Khinchin
theorem~\cite{LeadbetterLindgrenRootzen:1983}. It does, however, determine
convergence rates~\cite{GyorgyiETAL:2008}. In this paper we choose $a_n$ and
$b_n$ so that all limiting PDFs $p(y)=\frac{dF(y)}{dy}$ are standardized with
zero mean (via the location parameter) and unit standard deviation (via the
scale parameter), where $y:=a_n z + b_n$. That is,
\begin{align}
  \int z \, dF^n(a_n z + b_n) &= 0 \label{E:mean} \\
  \int z^2 \, dF^n(a_n z + b_n) &= 1. \label{E:sd}
\end{align}

\subsection{Maximum relative to average\label{sec:mra}}

Labelling the maximum as $x^{max}_n$, the maximum relative to the
average is defined as $z := x_n^{max} - \frac{1}{n}\sum_{i=1}^{n}
x_i$. By the central limit theorem, the average of $n$ iid uniform
deviates approaches a Gaussian distribution centered at $1/2$ with
a width that shrinks as $\mathcal{O}(n^{-1/2})$. The choice of
$a_n = 1/n$ in Eq.~\eqref{E:Weibull} shows that the maximum
approaches $1$ as an exponential distribution with a width that
shrinks as $\mathcal{O}(n^{-1})$. The scaling of $a_n$ is, to
leading order, generic up to
prefactors~\cite{LeadbetterLindgrenRootzen:1983}. Therefore, as
$n\to\infty$, the spread in the average dominates, and the PDF of
the maximum relative to the average is
\begin{equation}
  p(y) = \frac{1}{\sqrt{2\pi}}e^{-y^2/2},\label{E:mrha_rescaled}
\end{equation}
with zero mean and unit standard deviation.

\subsection{Maximum relative to initial value\label{sec:mri}}

Given an initial value $x_1$, the maximum relative to the initial
value is defined as $z := x^{max}_n - x_1$. The initial
value is a random variable with a distribution having a width of
$\mathcal{O}(1)$. Therefore, by the same argument as above, only
the spread in the initial value is observed for $n\to\infty$.
Specifically, two situations must be distinguished. With
probability $1/n$, the initial value is itself the maximum,
yielding a zero maximum relative to the initial value. Thus the
distribution is composed of a point mass of weight $1/n$ at $z=0$.
For $z>0$, the probability distribution is continuous and the
calculation of the probability density is as follows: without loss
of generality, label from among the remaining $(n-1)$ random
variables the maximum as the final one. Then
\begin{align}
  p(z) &= (n-1)\int_{0}^{x_n} dx_1
  \cdots \int_{0}^{x_n} dx_{n-1}
  \int_{0}^{1} dx_n  \notag \\
  &\times \prod_{i=1}^n \theta(x_i)\theta(1-x_i)
  \delta(z-(x_n-x_1)) \notag \\
  &= 1 - z^{n-1}, \label{E:mrhi}
\end{align}
where $(n-1)$ is a combinatorial factor taking into account all possible
locations of the maximum among the remaining $(n-1)$ random variables.

Substituting the PDF in Eq.~\eqref{E:mrhi} into
Eqs.~\eqref{E:mean},\eqref{E:sd}, the location and scale
parameters that maintain zero mean and unit standard deviation are
\begin{align}
  b_n &= \frac{1}{2}\frac{n}{(n+1)} = \frac{1}{2} - \frac{1}{2n} + 
  \mathcal{O}(n^{-2}) \\
  a_n &= \frac{\sqrt{3}}{6} - \frac{1}{4}\frac{\sqrt{3}}{n} +
  \mathcal{O}(n^{-2}).
\end{align}
Thus, to leading order, the location and scale parameters are
simply the mean and standard deviation of the uniform
distribution.  As $n\to\infty$ the rescaled PDF of the maximum
relative to the initial value converges to
\begin{equation}
  p(y) = \begin{cases} \sqrt{3}/6, &\quad -3/\sqrt{3} \le y \le 3/\sqrt{3} \\
                                    0, &\quad \text{otherwise.}\label{E:mrhi_rescaled}
              \end{cases}
\end{equation}
For more general cases, see~\cite{BurkhardtETAL:2007}. 

\subsection{Maximum relative to minimum\label{sec:mrm}}

Without loss of generality, labelling the minimum and the maximum
as the first and last random variables, respectively, the
distribution of the maximum relative to the minimum, $z :=
x_n-x_1$ is given by
\begin{align}
  p(z) &= n(n-1)\int_0^{x_n} dx_1 \int_{x_1}^{x_n} dx_2 \cdots
  \int_{x_1}^{x_n} dx_{n-1} \int_0^1 dx_n \notag \\
  &\times \prod_{i}^{n} \theta(x_i)\theta(1-x_i) \delta(z-(x_n-x_1)) \notag \\
  &= n(n-1)z^{(n-2)}(1-z),\label{E:maxmin}
\end{align}
where $n(n-1)$ is a combinatorial factor taking into account all possible
locations of the maximum and the minimum among the $n$ variables.

Substituting the PDF in Eq.~(\ref{E:maxmin}) into
Eqs.~\eqref{E:mean},\eqref{E:sd} the location and scale parameters
that maintain zero mean and unit standard deviation are
\begin{align}
  b_n &= \frac{n-1}{n+1} = 1 - \frac{2}{n} + \mathcal{O}(n^{-2}) \\
  a_n &= \frac{\sqrt{2}}{n} + \mathcal{O}(n^{-2}).
\end{align}
To leading order, the location parameter reflects the fact that
the distribution centers at $z = 1$ from below as $n\to\infty$.
Meanwhile, the scale parameter reflects the fact that the width of
the distribution shrinks as $\mathcal{O}(n^{-1})$ to leading
order. As $n\to\infty$ the rescaled PDF of the maximum relative to
the initial value converges to
\begin{equation}
  p(y) = \sqrt{2} (2-\sqrt{2}y) e^{\sqrt{2}y-2}, \quad -\infty < y \le
  \sqrt{2},
\label{E:maxmin_rescaled}
\end{equation}
and zero otherwise.

\subsection{Other distributions belonging to the Weibull class}

While the results given in Sections \ref{sec:mra}--\ref{sec:mrm} are for the
uniform distribution, the arguments used in their derivation can be easily
generalized to any other distribution. Namely, the widths under rescaling of
the maximum and the reference value should be compared in order to determine
the final limiting distribution for the extremal statistic. Generally, either
one or the other will dominate, apart from the special case when both widths
are rescaled in the same way, in which case there will be a convolution of the
two distributions (assuming that the $n$-independent prefactors are comparable
in magnitude). This is in particular true for another prominent example in the
Weibull class, the beta distribution, with PDF
\begin{equation}\label{beta}
  p(z) = \frac{z^{\gamma-1}(1-z)^{\delta-1}}{B(\gamma,\delta)}, \quad 0 \le z \le
  1,
\end{equation}
and zero otherwise, where $\gamma,\delta > 0$ and $B(\gamma,\delta)$
is the beta function. The maxima of the beta distribution are
Weibull-distributed, provided the width is rescaled with $a_n \sim
n^{-1/\delta}$. Thus, for example, the maximum measured with respect to the
average yields a Gaussian, a convolution, and a Weibull distribution for
$\delta < 2$, $\delta = 2$ and $\delta > 2$, respectively.

\section{Extremal statistics for stationary long-range correlated signals\label{S:stationary}}

The stationary long-range correlated series $X_i$ of length $n$ that we study
have a uniform underlying PDF on $[0,1]$ and two-time correlation
function that decays as
\begin{equation}
  C_{ij} = \langle X_i X_j \rangle - \langle X_i \rangle \langle X_j \rangle
  \sim |i-j|^{-(1-\alpha)},
\end{equation}
where the average corresponds to an ensemble average. Stationarity
requires that $\alpha < 1$ and long-range correlations are those
for which $0 < \alpha < 1$. By the Wiener-Khinchin theorem, the
Fourier transform of the two-time correlation function is simply
the power spectrum and both are equivalent descriptions. For
stationary long-range correlated series, it follows that the power
spectrum decays as $1/f^{\alpha}$.

In order to generate series of long-range correlated uniform deviates, we
employ the algorithm of Schreiber and
Schmitz~\cite{SchreiberSchmitz:1996,SchreiberSchmitz:2000}. The method works
iteratively to enforce a desired power spectrum by permuting iid random
variables drawn from a desired distribution. Each iteration consists of two
steps: adjusting the power spectrum of the random variables in Fourier space
with the appropriate filter, and then rank-order exchanging the
reversed-transformed variables with the original iid random variables. Thus,
the algorithm does not change the values of the variables drawn initially, but
does change their order. For the uniform distribution, only a small number of
iterations is required for the power spectrum to converge to the desired form
-- in our case, a power-law with slope $-\alpha$. More details of the method
are outlined in the Appendix.

The statistics for the extremal quantities are collected in
segments. Typically, we generated series of length $2^{19}$ to
$2^{22}$, cut up $2^7$ to $2^{10}$ times to produce blocks of
length $2^{11}$ to $2^{15}$. The process is repeated for a number
of generated series to produce histograms with
$\mathcal{O}(10^6-10^7)$ points.  The histograms are normalized
and rescaled to give $p(y)$, with zero mean and unit standard
deviation.

As mentioned in the introduction, the three classes of extreme distributions
for the maxima of iid random variables remain robust for a wide class of
dependent series~\cite{LeadbetterLindgrenRootzen:1983}, including, for
example, long-range correlated Gaussian series~\cite{Berman:1964}.
Numerically, we find that this is also true for the long-range correlated
uniform processes defined above. This suggests that the relevant mixing
conditions are satisfied for the case of long-range correlated uniform
deviates, although we have not checked this analytically~\footnote{Indeed, the
decay of the correlations is the same as in Gaussian series, for which a proof
exists.}.  Moreover, while there are strong finite size corrections to the
variation of the scale parameter $a_n$ with $n$, even the asymptotic behavior
or leading order scaling is identical to the iid case, $a_n =1/n$. Our
numerics indicate that for small system sizes the higher order corrections are
positive.  Such a behavior is not generally expected since, owing to
correlations, the leading order scaling of $a_n$ may be different than those
in the iid case~\cite{LeadbetterLindgrenRootzen:1983}.  In the following, we
show that analogous results hold for the three relative maxima, i.e., the
distributions derived in
Eqs.~\eqref{E:mrha_rescaled},\eqref{E:mrhi_rescaled},\eqref{E:maxmin_rescaled}
for iid random variables also apply to the dependent case.

\subsection{Maximum relative to average}

\begin{figure}
\includegraphics{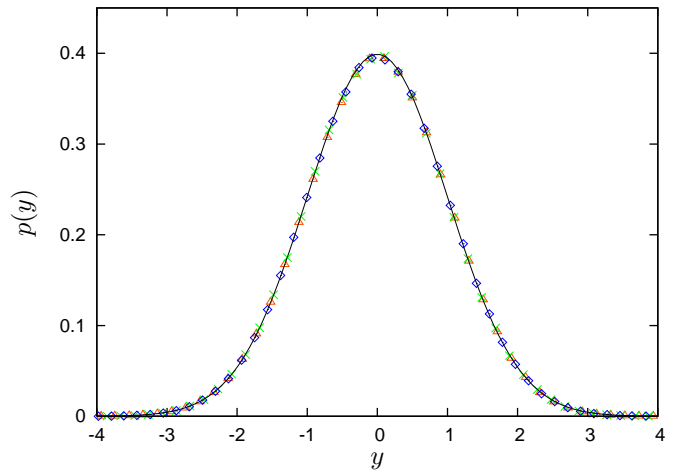}
\caption{(Color online) Distribution of maximum relative to average for
  sequences of length $2^{11},2^{13},2^{15}$ (triangles, crosses, diamonds)
  and $\alpha = 0.5$, with zero mean and unit standard deviation. A Gaussian
  is plotted alongside as a solid line.\label{F:mrha_a0.5}}
\end{figure}

Fig.~\ref{F:mrha_a0.5} plots the maximum relative to the average for blocks of
length $2^{11},2^{13},2^{15}$ (triangles, crosses, diamonds) and $\alpha =
0.5$. The curves are barely distinguishable from the iid limiting Gaussian
distribution, and the convergence is rapid for the system sizes
examined. Numerically, we find that the same holds for $0 \le \alpha < 1$ (not
shown). These results are expected since the width of the limiting Gaussian
distribution of the average decays as
$n^{(\alpha-1)/2}$~\cite{BouchaudGeorges:1990,IbragimovLinnik:1971} while the
width of the distribution of the maximum decays asymptotically as $n^{-1}$ as
mentioned above. Thus, the average dominates for any $\alpha$ implying that a
Gaussian distribution is observed asymptotically.

\subsection{Maximum relative to initial value}

\begin{figure}
\includegraphics{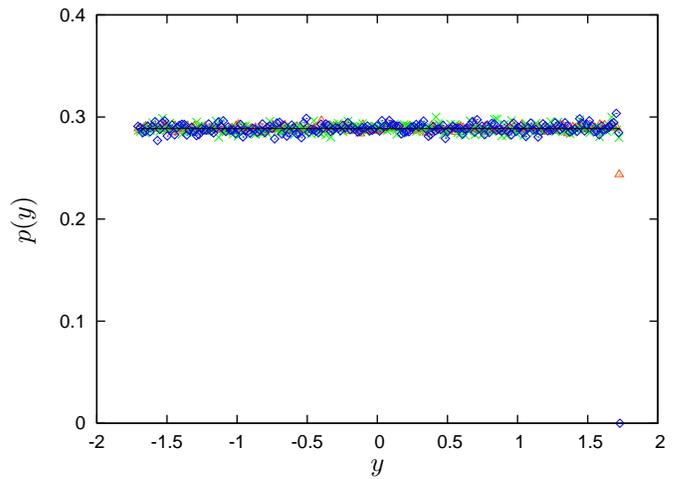}
\caption{(Color online) Distribution of maximum relative to initial value for
  sequences of length $2^{11},2^{13},2^{15}$ (triangles, crosses, diamonds)
  and $\alpha = 0.5$, with zero mean and unit standard deviation. A uniform
  distribution is plotted alongside as a solid line.\label{F:mrhi_a0.5}}
\end{figure}

\begin{figure}
\includegraphics{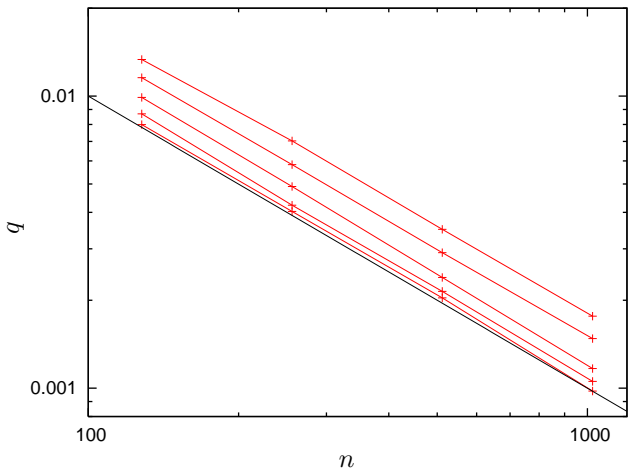}
\caption{(Color online) Decay in the fraction of zero maxima relative to
  initial value with increasing system size, for $\alpha =
  0.95,0.8,0.6,0.4,0.2$ (top to bottom). The decay $1/n$ for iid series is
  shown by the solid black line.\label{F:zero_scaling}}
\end{figure}

Fig.~\ref{F:mrhi_a0.5} plots the maximum relative to the initial
value for blocks of length $2^{11},2^{13},2^{15}$ (triangles,
crosses, diamonds) and $\alpha = 0.5$. The curves closely follow a
straight horizontal line of height $\sqrt{3}/6$. We find the same
behavior for other $\alpha$ with $0 \leq \alpha < 1$ (not shown).
If we assume that the distribution approaches the same limit as in
the iid case, then the exact form for finite $n$ is given by
Eq.~\eqref{E:mrhi}. The function approaches zero at its upper
endpoint very rapidly, and this feature is also observed in
Fig.~\ref{F:mrhi_a0.5} for the dependent case.
Since the distribution consists of a point mass at zero maximum
relative to the initial value, it is particularly convenient to
analyze this behavior with increasing $\alpha$.
Fig.~\ref{F:zero_scaling} demonstrates that the fraction $p$ of
times when the initial value is also the maximum decreases as
$1/n$.
However, the amplitude of the decay increases with $\alpha$. An
intuitive explanation for this behavior is as follows: persistence
along the series increases with $\alpha$, i.e. the series has
memory and is more likely to continue in the same direction as
previously. A series that starts with a downward trend is more
likely to produce a zero maximum relative to the initial value.
Such persistent trajectories are more prevalent with increasing
$\alpha$, and therefore the amplitude of the point mass is
enhanced relative to the iid case.

\begin{figure}
\includegraphics*[width=\columnwidth]{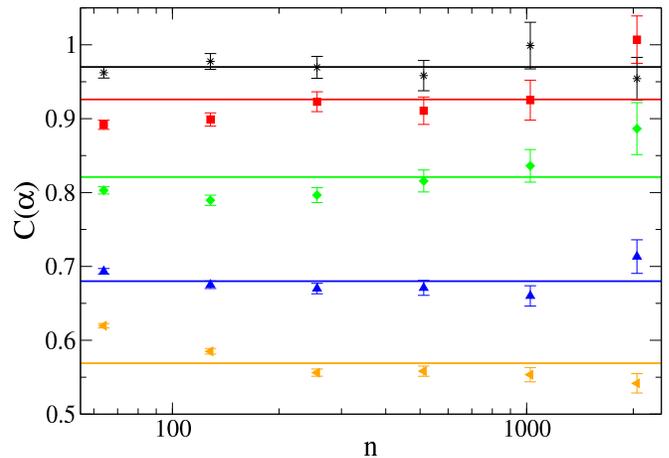}
\caption{(Color online) Measurements of $C(\alpha)$ based on
  Eq.~\eqref{E:c_alpha} for $\alpha = 0.2,0.4,0.6,0.8,0.95$ (top to bottom)
  and $n=64,128,256,512,1024,2048$. For each $\alpha$, the lines are
  averages over the six system sizes. Error bars are estimates of statistical
  error (via standard error propagation) and represent one standard deviation:
  over $N$ measurements the number of instances of the maximum also being the
  initial value is binomially distributed with expectation $Np$ and variance
  $Np(1-p)$, where $p=1/n$ is the probability that the maximum lies at the
  initial value.
\label{F:c_alpha}}
\end{figure}

In the presence of correlations, convergence to limiting distributions is
typically slower than in the iid case \cite{Coles:2001}.  The effect of
increasing $\alpha$ (i.e. correlations) can be taken into account by replacing
the block size $n$ by a smaller size $n_{\text{eff}} = C(\alpha) n$, where
$C(\alpha) < 1$ is an $\alpha$-dependent constant. If we assume a similar
behavior for the maximum measured relative to the minimum, then the simplest
modification to Eq.~(\ref{E:mrhi}) for correlated series would consist of a
point mass with weight $1/(C(\alpha)n)$ at zero, together with a density
\begin{equation}
  p(z) = 1 - z^{C(\alpha)n-1} \label{E:c_alpha}
\end{equation}
for $z > 0$. Fig.~\ref{F:c_alpha} plots estimates of $C(\alpha)$ for various
$\alpha$, suggesting that a decreased effective degrees of freedom description
is valid.

\subsection{Maximum relative to minimum}

\begin{figure}
\includegraphics{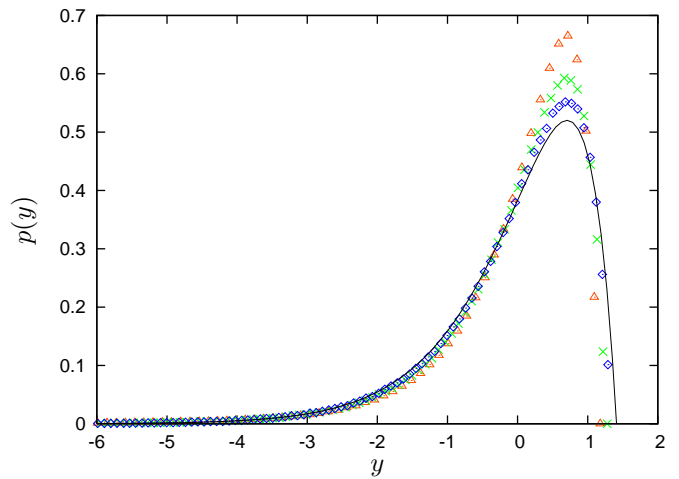}
\caption{(Color online) Distribution of maximum relative to minimum for
  sequences of length $2^{11},2^{13},2^{15}$ (triangles, crosses, diamonds)
  and $\alpha = 0.5$, with zero mean and unit standard deviation. The
  distribution in Eq.~\eqref{E:maxmin_rescaled} is plotted alongside as a
  solid line.\label{F:maxmin_a0.5}}
\end{figure}

Fig.~\ref{F:maxmin_a0.5} plots the maximum relative to the minimum for blocks
of length $2^{11},2^{13},2^{15}$ (red, green, blue) and $\alpha =
0.5$. Convergence is evidently slower for this maximal quantity and worsens
with increasing $\alpha$ (not shown). We speculate that this is because the
maximum and minimum approach their respective limiting distributions at the
same rate, giving rise to the convoluted distribution derived in
Eq.~(\ref{E:maxmin}) in the iid case. 

\subsection{Other distributions belonging to the Weibull class}

For long-range correlated processes obeying the beta distribution defined in
Eq.~(\ref{beta}), the same picture emerges as above once the relative
rescalings of the maximum and the respective reference value are taken into
account.  Our numerics~\cite{numerics} indicate that the distribution of the
maxima and the minima both approach a Weibull distribution and that $a_n$
scales asymptotically as $n^{-1/\delta}$ for the maximum and $n^{-1/\gamma}$
for the minimum --- as in the iid case. As in the case of the uniform
distribution, the width of the limiting Gaussian distribution of the average
scales as $n^{(\alpha-1)/2}$. Thus, depending on the particular choice of
$\alpha$, $\gamma$ and $\delta$, different limiting distributions are
obtained. For example, for $\gamma = \delta = 2$ the extremal statistics obey
respectively a Gaussian, a beta distribution with $\gamma = \delta = 2$, and a
convoluted distribution for the maximum measured with respect to the average,
initial value, and minimum.
 
In general, if a given Weibull process satisfies the relevant mixing
conditions~\cite{LeadbetterLindgrenRootzen:1983} and $a_n$ scales
asymptotically like its uncorrelated counterpart, then, barring slow
convergence, one of three scenarios will occur: (a) the width of the maximum
distribution shrinks slower than that of the reference, in which case the
maximum distribution is observed, (b) the width of the reference distribution
shrinks slower than that of the maximum, in which case the reference
distribution is observed, (c) the widths of the maximum and reference
distributions scale at the same rate and are of comparable amplitude, in which
case a non-trivial convolution is observed.

\section{Return times\label{S:returns}}

Return times are useful indicators for analyzing time series and are
particularly relevant when forecasting extreme events, e.g.  floods or large
earthquakes. For example, a flood levee may be constructed so as to permit
flooding only once every $10,000$ years, on average. Supposing that at time
$t_i$ an event of magnitude $x_i$ exceeds some threshold $q$, and that the
threshold is subsequently exceeded for the first time by an event $x_j$ at
$t_j$, then the return time is defined as $t := t_j-t_i$. Since in our context
time marches in unit steps, the smallest return time is $1$, i.e. as a result
of two consecutive events that exceed the threshold.

In our simulations we generate series of length $2^{19},\ldots,2^{22}$ a
repeated number of times ($\sim 10^2$ to $10^5$). The return times derived
from these series are then combined into a histogram, from which we construct
the normalized return time distribution $p(t)$.

\begin{figure*}
\includegraphics{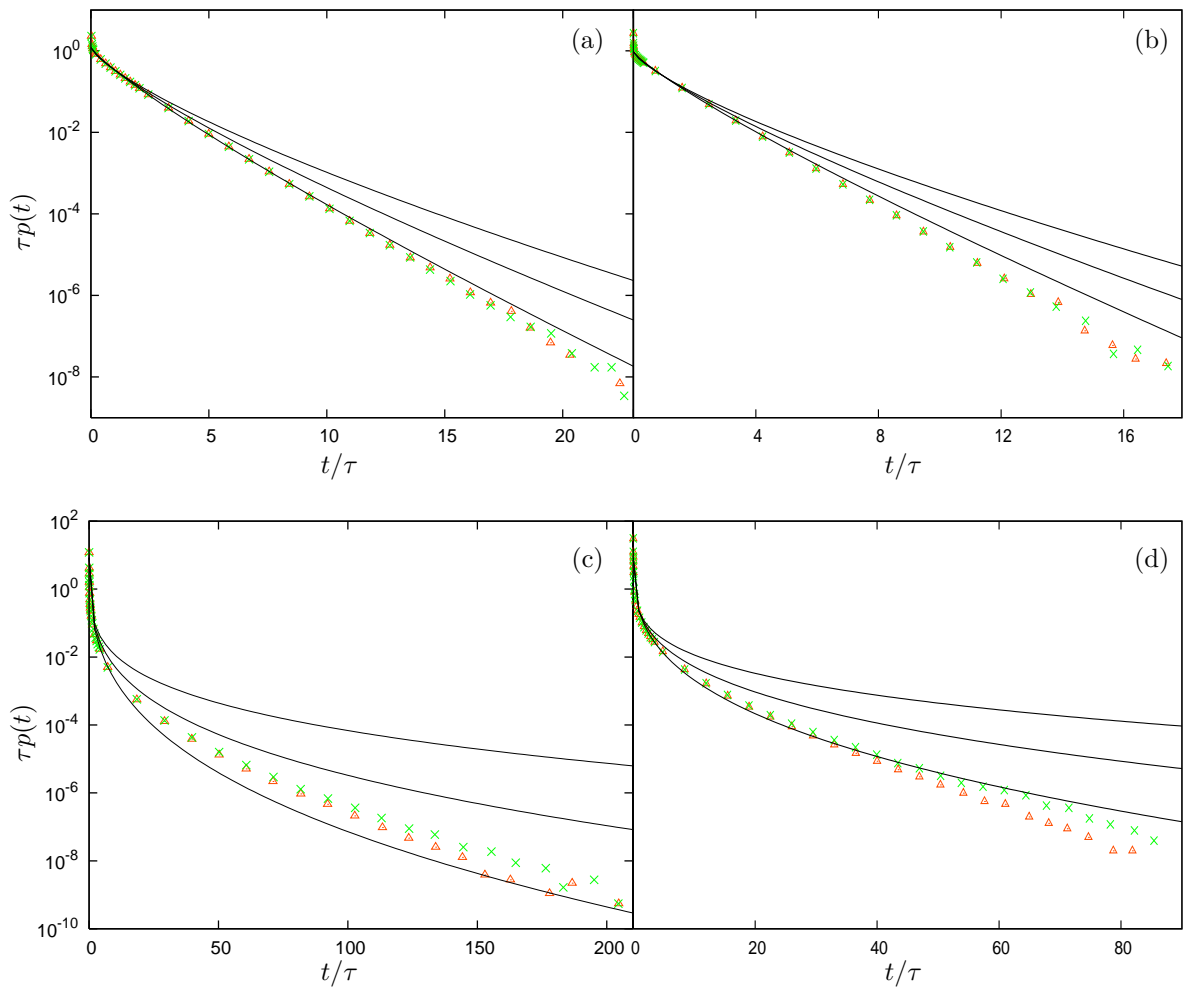}
\caption{(Color online) Top panels: distribution of return times above the (a)
  $97.9\%$, (b) $99.7\%$ quantile for sequences of length $2^{19},2^{22}$
  (triangles, crosses) and $\alpha = 0.2$, with mean return times (a) $\tau
  \approx 48$, (b) $\tau \approx 287$. Stretched exponential ansatzes
  (see Eq.~\eqref{E:stretched}) are plotted alongside for $\alpha = 0.25,0.2,0.15$
  from top to bottom. Bottom panels: distribution of return times above the
  (c) $97.9\%$, (d) $99.7\%$ quantile for sequences of length $2^{19},2^{21}$
  (triangles, crosses) and $\alpha = 0.7$, with mean return times (c) $\tau
  \approx 48$, (d) $\tau \approx 287$. Stretched exponential ansatzes are
  plotted alongside for $\alpha = 0.75,0.7,0.65$ from top to
  bottom. \label{F:ret_4_panels}}
\end{figure*}

The iid scenario is described by a Poisson point process, giving exponentially
distributed return times. For stationary long-range correlated series,
meanwhile, the authors of~\cite{EichnerETAL:2007,AltmannKantz:2005} proposed
the following fit-free (but nevertheless $\alpha$-dependent) ansatz for the
return time distribution:
\begin{equation}
  \tau p(t) = a_{\alpha} \exp{[-b_{\alpha} (t/\tau)^{1-\alpha}]}, \label{E:stretched}
\end{equation}
where $\tau$ is the mean return time, and $a_{\alpha}$ and $b_{\alpha}$ are
$\alpha$-dependent constants of normalization fixed by
\begin{align}
  \sum_{t=1}^{\infty} p(t) &= 1, \\
  \sum_{t=1}^{\infty} t p(t) &= \tau.
\end{align}
We test this stretched exponential ansatz for long-range correlated uniform
deviates.  Fig.~\ref{F:ret_4_panels}(a),(b) plots numerical results for
$\alpha = 0.2$ and two different quantiles. Stretched exponential curves are
drawn alongside for $\alpha = 0.25,0.2,0.15$ from top to bottom. In
Fig.~\ref{F:ret_4_panels}(a) the errors are within the symbol size up to
$t/\tau \approx 16$, with larger statistical fluctuations thereafter. For
Fig.~\ref{F:ret_4_panels}(b), errors remain within symbol size up to $t/\tau
\approx 13$. The agreement between the numerical results and the ansatz is
poor.  Increasing the quantile worsens the agreement. For the results shown,
there appears to be no systematic trend towards the ansatz with increasing
system size.  Figs.~\ref{F:ret_4_panels}(c),(d) plot numerical results for
$\alpha = 0.7$ and two different quantiles. Stretched exponential curves are
drawn alongside for $\alpha = 0.75,0.7,0.65$ from top to bottom. In
Fig.~\ref{F:ret_4_panels}(c) the errors are within the symbol size up to
$t/\tau \approx 140$, with larger statistical fluctuations thereafter. For
Fig.~\ref{F:ret_4_panels}(d), errors remain within symbol size up to $t/\tau
\approx 50$. The agreement between the numerical results and the ansatz is
better than at $\alpha = 0.2$, but still unsatisfactory. Increasing the
quantile pushes the numerical results further away from the $\alpha = 0.7$
ansatz, although there does seem to be an improvement in the overall
shape. Increasing the system size does reveal a trend towards the ansatz curve
for $\alpha = 0.7$, although much larger system sizes would be required to
examine whether this trend converges. A similar picture is obtained for the
beta distribution~\cite{numerics}, i.e. poor agreement between numerics and
ansatz that worsens with increasing quantile.

\begin{figure}
\includegraphics{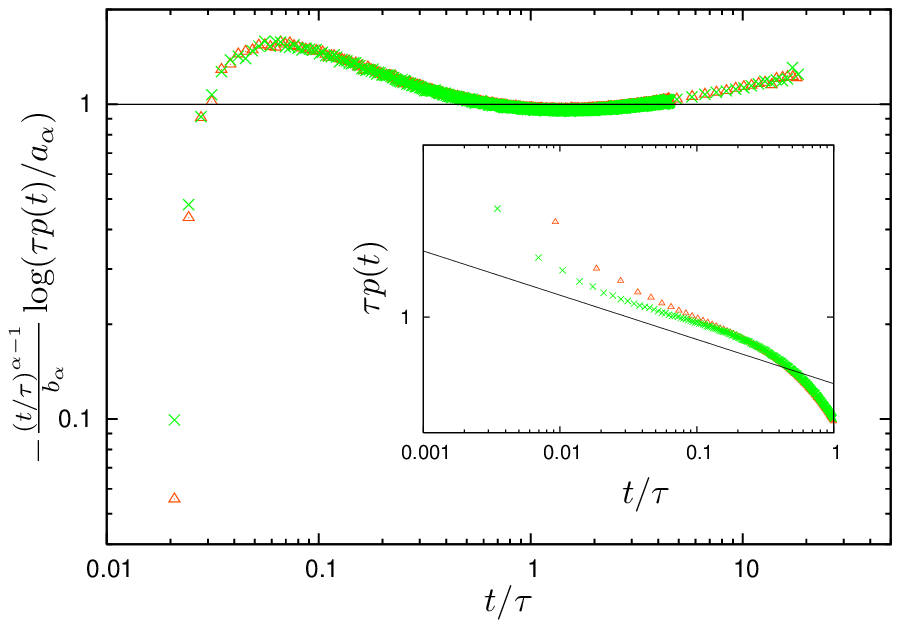}
\caption{(Color online) Main plot: distribution of return times above the
  $99.7\%$ quantile for sequences of lengths $2^{19},2^{21}$ (triangles,
  crosses) for $\alpha = 0.2$. Inset: distribution of return times above the
  $99.1\%,99.7\%$ quantiles (triangles, crosses) for sequences of length
  $2^{21}$ and $\alpha = 0.2$. A power law with slope $-\alpha$ is plotted
  alongside.\label{F:ret_asymp_a0.2}}
\end{figure}

\begin{figure}
\includegraphics{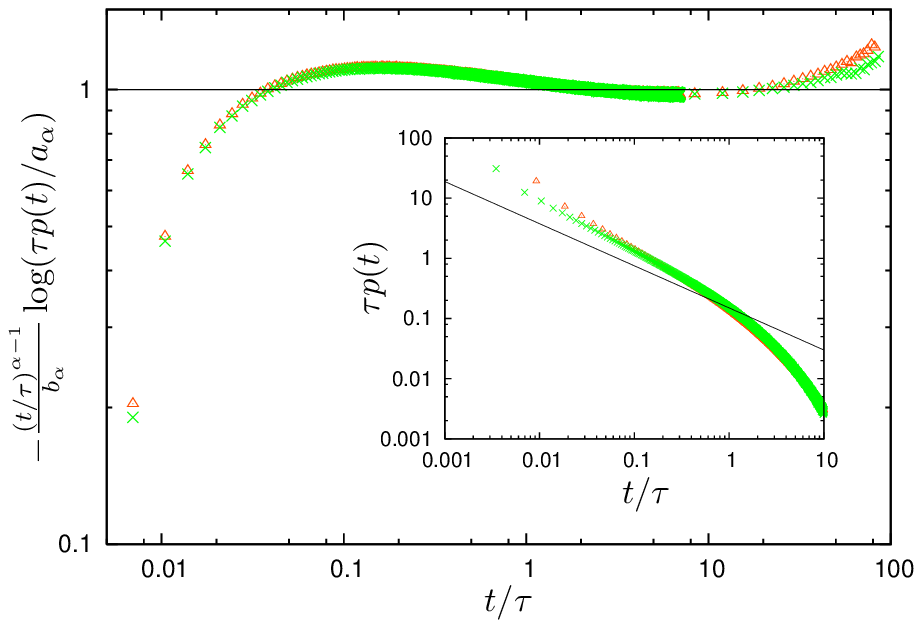}
\caption{(Color online) Main plot: distribution of return times above the
  $99.7\%$ quantile for sequences of lengths $2^{19},2^{21}$ (triangles,
  crosses) for $\alpha = 0.7$. Inset: distribution of return times above the
  $99.1\%,99.7\%$ quantiles (triangles, crosses) for sequences of length
  $2^{21}$ and $\alpha = 0.7$. A power law with slope $-\alpha$ is plotted
  alongside. \label{F:ret_asymp_a0.7}}
\end{figure}

These observations are confirmed by plotting
$-\frac{{t/\tau}^{\alpha-1}}{b_{\alpha}} \log(\tau p(t)/a_{\alpha})$ against
$t/\tau$. If the proposed form of a stretched exponential is correct, then the
numerical curves should approach a constant
asymptotically~\cite{AltmannKantz:2005}. The main plots in
Figs.~\ref{F:ret_asymp_a0.2},\ref{F:ret_asymp_a0.7} show the results for
$\alpha = 0.2,0.7$. The numerical curves bend around the horizontal line,
suggesting that a stretched exponential is too simplistic to describe the
functional form of the asymptote. We find that the agreement with the ansatz
improves with increasing $\alpha$. For the most part, the dependence on series
length is weak for the series we considered.

The particularly strong deviations from the proposed stretched exponential
given in Eq.~(\ref{E:stretched}) for $t < \tau$ have also been observed for
other stationary long-range correlated series~\cite{EichnerETAL:2007}.  The
authors conjecture that instead the initial asymptote rather follows a power
law with slope $-\alpha$. We test this hypothesis in the insets of
Figs.~\ref{F:ret_asymp_a0.2},\ref{F:ret_asymp_a0.7} for $\alpha = 0.2,0.7$
respectively. The agreement with the power-law ansatz is reasonable and
improves with decreasing $\alpha$ and increasing quantile.

To summarize, while it is still possible that the ansatz for the return time
distribution of stationary long-range correlated series given in
Eq.~(\ref{E:stretched}) generally holds for $t>\tau$ in the limit of
infinitely long time series, there are significant deviations from it for the
long but finite time series considered here. These deviations also depend on
the threshold in a counter-intuitive way such that the deviations are stronger
for higher thresholds.

\section{Conclusion\label{S:conclusion}}

Long-range correlated series with an underlying distribution belonging to the
Weibull class of extreme values have received little attention in the
literature. We believe that our study will be of relevance to certain
quantities that approach finite bounds sufficiently slowly (for a precise
mathematical statement, see~\cite{deHaanFerreira:2006}). Possible examples
include humidity~\cite{NREL:1993}, strength of materials~\cite{Weibull:1951},
and critical path analysis~\cite{Thornley:2001}. In the latter case, the beta
distribution is often used for modelling completion times of activities.

While in the mathematical literature one can find important extensions of iid
extreme value statistics to stationary dependent
series~\cite{LeadbetterLindgrenRootzen:1983,LeadbetterRootzen:1988}, we have
focussed on extremal quantities that are measured with respect to a reference
point that is itself a random variable. Apart from the mathematical interest,
this is motivated by practical applications. To pick just one of the examples,
the maximum relative to the minimum gives a measure of the full range
traversed by a trajectory in a stochastic process. In many cases, the
distribution of the reference point dominates or convolves with the
distribution of the extremes giving rise to extremal distributions very
different from the Weibull distribution. Moreover, we found that the extremal
distributions in the correlated stationary series converge to their iid
counterparts for the specific processes considered.

While it was proposed that the form of return interval distributions in
correlated stationary series asymptotically approaches a stretched exponential
for large return intervals independent of the underlying
distribution~\cite{EichnerETAL:2007}, our results for the uniform and beta
distribution do not support this.  Although a stretched exponential is
appealing and fully-determined once $\alpha$ has been estimated, we believe
that this can only be a first approximation.  However, a power-law decay with
slope $-\alpha$ for short return intervals is more convincing. Together with
the results presented in Ref.~\cite{EichnerETAL:2007}, this suggests that this
behavior might be universal and independent of the underlying distribution.

\begin{acknowledgments}
The authors would like to thank G\'eza Gy\"orgyi and Holger Kantz for useful
discussions.
\end{acknowledgments}

\section{Appendix}

The numerical results presented in this article are based on an algorithm
proposed in Refs.~\cite{SchreiberSchmitz:1996,SchreiberSchmitz:2000} to
generate realizations of stochastic processes with a desired PDF and desired
correlations. While the algorithm has been extensively tested and is
well-established, questions remain as to the nature of the sample paths, the
step size distribution, etc.  To investigate these points, we apply the
algorithm to the case of a uniform PDF with $\alpha = 2$, which should
correspond to a diffusion-like process. This is indeed what we find. 

\begin{figure}
\includegraphics{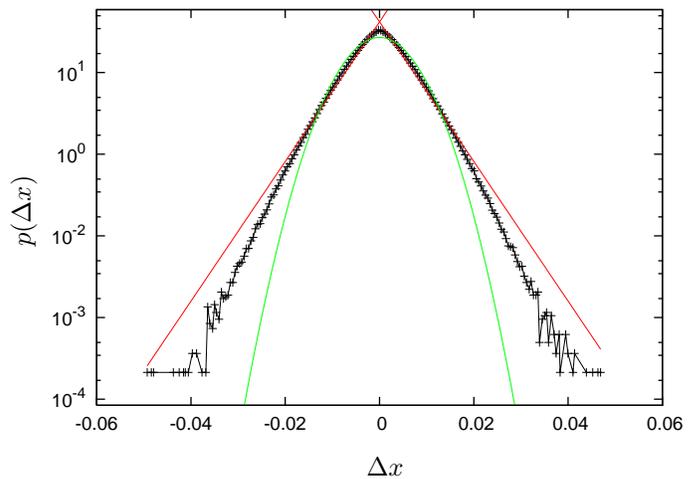}
\caption{(Color online) Distribution of step sizes for series of length
  $32768$ and $\alpha = 2$. A Gaussian and a symmetric exponential, both with
  the same standard deviation as the data, have been plotted alongside as
  solid lines.\label{F:step_sizes}}
\end{figure}

In Fig.~\ref{F:step_sizes} we plot the distribution of the step sizes between
successive points in series of length $32768$ and $\alpha = 2$. The
distribution lies somewhere between a Gaussian and a symmetric exponential,
i.e. distributions with suppressed tails (as opposed to, say, power-law tails
arising from an underlying L\'{e}vy motion).

\begin{figure}
\includegraphics{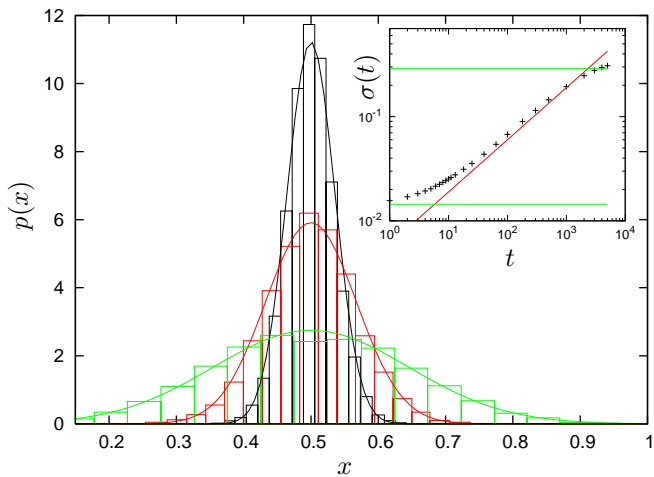}
\caption{(Color online) Time evolution of the PDF for sequences initially
starting within $0.475 \le x \le 0.525$ after $25,100,500$ steps (top to
bottom), for $L = 32768$ and $\alpha = 2$. Gaussians with the same standard
deviation as the data are plotted alongside. Inset: Time evolution of the
standard deviation of the PDF (crosses) for sequences initially starting
within $0.475 \le x \le 0.525$. The standard deviations of the initial and
final uniform distributions are $\sqrt{3}/120$ and $\sqrt{3}/6$ (horizontal
lines), respectively. For a range of intermediate times the standard deviation
grows with the square root of the number of steps (diagonal
line).\label{F:pdf_evol}}

\end{figure}

To minimize the effect of the finite support of the uniform distribution, we
also studied the time evolution of those realizations or series beginning in
the middle of the interval. The inset of Fig.~\ref{F:pdf_evol} plots the
standard deviation of the distribution of these series as a function of
time. Numerically, among our ensemble of series we included all those that
began with values in the narrow interval $0.475 \le x \le 0.525$. Since the
random variables are drawn from a uniform distribution, at time $t=0$ their
distribution has standard deviation $\sigma(t = 0) = \sqrt{3}/120$, which is
indicated by the lower horizontal line. For intermediate times the
distribution spreads with a standard deviation $\sigma(t) \propto
\sqrt{t}$. For long times the distribution crosses over to a uniform spread
across the interval, as indicated by the upper horizontal line with standard
deviation $\sigma(t\to\infty) = \sqrt{3}/6$. This is consistent with
diffusion. However, an idiosynchracy of the Schreiber-Schmitz method is that
the possible values encountered in the series are quenched at time
$t=0$. Therefore, preselected realizations starting in the middle narrow
interval exhaust a fraction of the uniform deviates from this region of the
$[0,1]$ interval and, correspondingly, a relatively larger amount of values is
subsequently encountered around the endpoints. This is the reason for the
overshoot of the standard deviation past the upper horizontal line, since the
distribution is slightly enhanced at its endpoints (and slightly diminished
around $0.5$) with respect to a uniform distribution. But after all the values
in the series have been accounted for, the distribution of the original
uniform draw is recovered. Fig.~\ref{F:pdf_evol} illustrates that, for
intermediate times, the evolving distributions are well approximated by
Gaussians, as expected for diffusion.

\bibliography{articles,books,j2}

\begin{thebibliography}{45}
\bibitem[{num()}]{numerics}
\bibinfo{note}{We simulated long-range correlated processes with $\alpha=0.5$
  based on the beta distribution for $\gamma = \delta = 2$, $\gamma = 2$,
  $\delta = 5$ and $\gamma = 2$, $\delta = 3/2$.}
\end{thebibliography}

\end{document}